\documentclass[fleqn,usenatbib]{mnras}

\usepackage{newtxtext,newtxmath}

\usepackage[T1]{fontenc}

\DeclareRobustCommand{\VAN}[3]{#2}
\let\VANthebibliography\thebibliography
\def\thebibliography{\DeclareRobustCommand{\VAN}[3]{##3}\VANthebibliography}

\usepackage{graphicx}	
\usepackage{amsmath}	

\usepackage{verbatim}
\usepackage[utf8]{inputenc}
\usepackage{hyperref}
\usepackage[dvipsnames]{xcolor}
\usepackage{units}
\usepackage{textcomp}
\usepackage{colortbl}

\definecolor{Berry}{HTML}{FF2052}

\usepackage{pgfplots}
\usepackage{filecontents}

\allowdisplaybreaks

\title[Dust processing at different stages in Cas A]{From total destruction to complete survival: Dust processing at different evolutionary stages in the supernova remnant Cassiopeia A}

\author[F. Kirchschlager et al.]{Florian Kirchschlager\href{https://orcid.org/0000-0002-3036-0184}{\includegraphics[trim=0cm -35cm 0cm 0cm, clip=true,width=0.27cm]{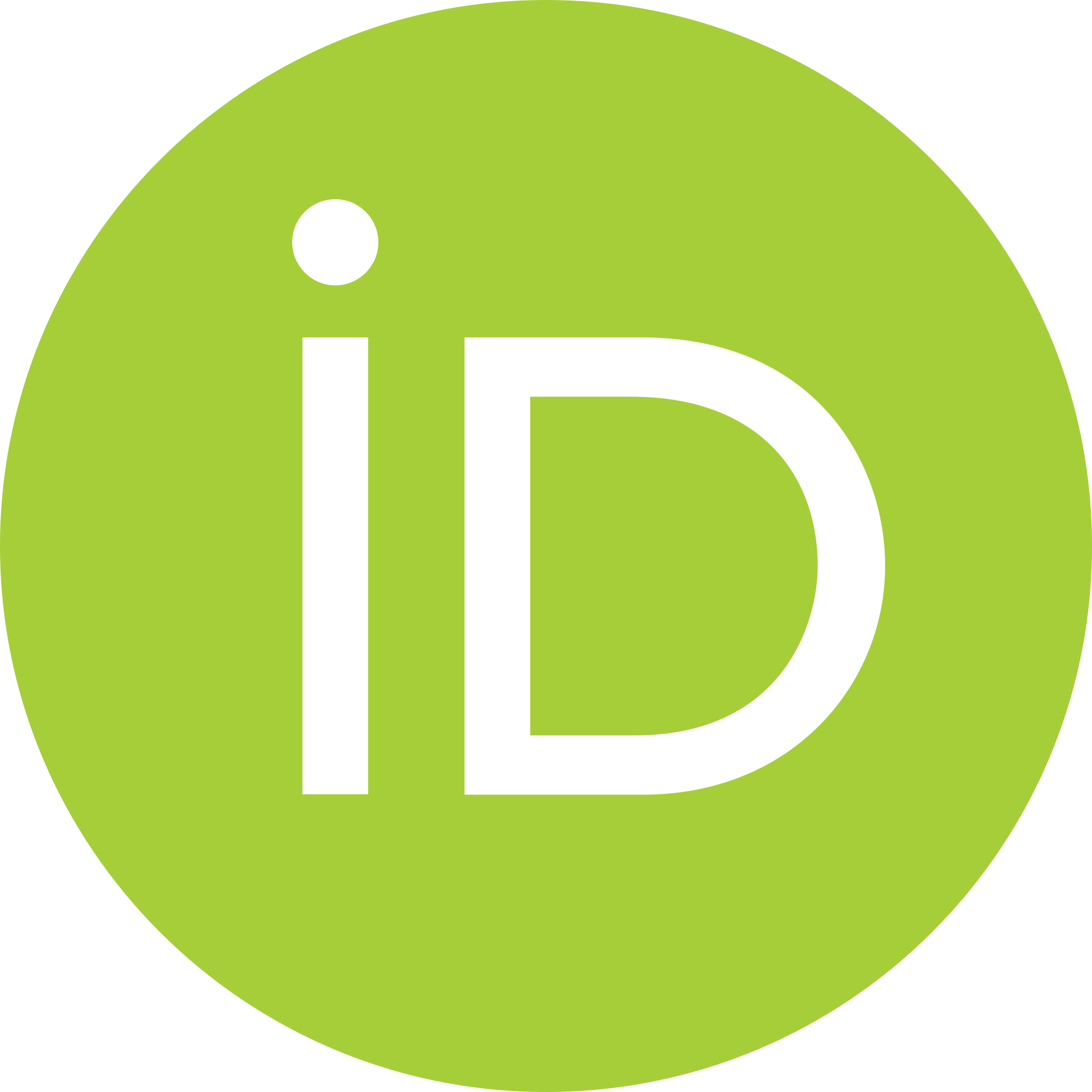}},$^{1}$\thanks{E-mail: florian.kirchschlager@ugent.be}
Nina S. Sartorio\href{http://orcid.org/0000-0003-2138-5192}{\includegraphics[trim=0cm -35cm 0cm 0cm, clip=true,width=0.27cm]{Pics/orcidID.jpg}},$^{1}$
Ilse De Looze\href{http://orcid.org/0000-0001-9419-6355}{\includegraphics[trim=0cm -35cm 0cm 0cm, clip=true,width=0.27cm]{Pics/orcidID.jpg}},$^{1}$
M. J. Barlow\href{http://orcid.org/0000-0002-3875-1171}{\includegraphics[trim=0cm -35cm 0cm 0cm, clip=true,width=0.27cm]{Pics/orcidID.jpg}},$^{2}$\newauthor
Franziska D. Schmidt$^{2}$ and Felix D. Priestley$^{3}$ 
\\
$^{1}$Sterrenkundig Observatorium, Ghent University, Krijgslaan 281-S9, B9000 Gent, Belgium\\
$^{2}$Department of Physics and Astronomy, University College London, Gower Street, London WC1E 6BT, United Kingdom\\
$^{3}$School of Physics and Astronomy, Cardiff University, Queen’s Buildings, The Parade, Cardiff CF24 3AA, United Kingdom
}

\date{Accepted 2024 January 31. Received 2024 January 30; in original form 2023 September 01}

\pubyear{2023}

\begin{document}
\label{firstpage}
\pagerange{\pageref{firstpage}--\pageref{lastpage}}
\maketitle

\begin{abstract}
The expanding ejecta of supernova remnants (SNRs) are believed to form dust in dense clumps of gas. Before the dust can be expelled into the interstellar medium and contribute to the interstellar dust budget, it has to survive the reverse shock that is generated through the interaction of the preceding supernova blast wave with the surrounding medium. The conditions under which the reverse shock hits the clumps change with remnant age and define the dust survival rate. To study the dust destruction in the SNR Cassiopeia~A, we conduct magnetohydrodynamical simulations of the evolution of a supernova blast wave and of the reverse shock. In a second step we use these evolving conditions to model clumps that are disrupted by the reverse shock at different remnant ages. Finally, we compute the amount of dust that is destroyed by the impact of the reverse shock. We find that most of the dust in the SNR is hit by the reverse shock within the first $350\,$yr after the SN explosion. While the dust destruction in the first $200\,$yr is almost complete, we expect greater dust survival rates at later times and almost total survival for clumps that are first impacted at ages beyond $1000\,$yr. Integrated over the entire evolution of the SNR, the dust mass shows the lowest survival fraction (17 per cent) for the smallest grains (1$\,$nm) and the highest survival fraction (28 per cent) for the largest grains (1000$\,$nm). 
\end{abstract}

\begin{keywords}
ISM: supernova remnants -- shock waves -- supernovae: individual: Cassiopeia A -- supernovae: general -- magnetohydrodynamics -- dust, extinction\\[-0.45cm]  
\end{keywords}




\section{Introduction}
The formation of dust in the ejecta of core-collapse supernovae (CCSNe) has been detected in the last three decades by different methods, such as the steepening of the optical brightness decay due to increased dust extinction (e.g. \citealt{Lucy1989}), an infrared excess due to thermal dust emission (e.g. \citealt{Wooden1993, Wesson2015}), and a red-blue-asymmetry of emission lines due to higher dust absorption from the far side of the expanding ejecta (e.g. \citealt{Lucy1989, Bevan2016}). The dust grains are expected to be formed in dense clumps of gas that move with the expanding ejecta behind the forward shock (\citealt{Kotak2009}). At the edge of the supernova remnant (SNR), the forward shock wave is interacting with circumstellar material (CSM) and interstellar material (ISM) which are decelerating the forward shock and creating a second shock wave, the reverse shock. Initially directed outwards, the reverse shock reverts at later ages and moves toward the ejecta centre. At any phase, the velocity of the reverse shock is lower than the forward shock and the expanding ejecta. As a result the clumps that formed in the ejecta are inevitably on a collision course with the reverse shock and will be disrupted by the high-energy impact. The dust, formerly located in the clumps, is then exposed to harsh conditions in the shocked gas and can be (partly) destroyed by gas-grain interactions (thermal and kinetic sputtering; e.g. \citealt{Barlow1978}) or grain-grain collisions (fragmentation or vaporization; e.g. \citealt{Jones1996}). The impact of the reverse shock causes a grain-size dependent acceleration and thus displacement of the dust and a re-distribution of the grain size distribution (e.g. \citealt{Kirchschlager2019b}).

A number of previous studies have investigated dust destruction fractions caused by the passage of an SNR reverse shock. Their formalisms, approaches, and models vary widely, as do the resulting dust survival rates. \cite{Nozawa2007} found that depending on the SN explosion energy, between 0 and 80$\,$per~cent of the initial dust mass can survive. The survival rate derived by \cite{Bianchi2007} is between 2 and 20$\,$per~cent, depending on the density of the surrounding CSM. The wide spread of survival fractions is continued by \cite{Nath2008} (80 – 99$\,$per~cent), \cite{Silvia2010, Silvia2012} (0 – 99$\,$per~cent), \cite{Biscaro2016} (6 – 11$\,$per~cent), \cite{Bocchio2016} (1 – 8$\,$per~cent), \cite{Micelotta2016} (12 – 16$\,$per~cent), \cite{Kirchschlager2019b} (0 – 40$\,$per~cent), \cite{Slavin2020} (10 – 50$\,$per~cent), and \cite{Kirchschlager2023} (0 – 99$\,$per~cent). The different studies emphasize the strong dependence of the survival rate on initial dust properties such as grain size and material, on different destruction processes (grain-grain collisions are often ignored), but also on properties of the ejecta such as the shock velocity, gas densities in the clumps and in the surrounding CSM, explosion conditions as well as magnetic field strengths and orientations. The ejecta and shock properties are not constant over time but are subject to the remnant evolution. Consequently, the remnant age has a significant impact on the dust survival fractions.

After a few hundred to thousand years SNRs are several parsecs in size (\citealt{Bocchio2016, Micelotta2016,Martinez2019,Slavin2020, Kirchschlager2022, Kirchschlager2024a}). In order to model the remnant evolution in an MHD simulation, a suitably similar box size is required. Clumps in the SNR, on the other hand, are comparatively small -- the clump radii observed in the SNR Cassiopeia A (Cas A) are in the range $(0.5-2.5)\times 10^{16}\,$cm (\citealt{Fesen2011}), which requires a high resolution to model the interaction of the reverse shock with a clump. Recently, \cite{Kirchschlager2023} used a grid cell size of $10^{14}\,$cm (${\sim3}\times 10^{-5}\,$pc)  to model a single clump that is disrupted by the reverse shock. The simultaneous modelling of clump destruction and remnant evolution represents a major challenge, even in 2D, since a high computational effort is required. In previous studies in which the remnant evolution was taken into account, either no clumps were considered at all (e.g. \citealt{Bocchio2016}), clumps were modelled at low resolution (e.g.~${>}75\times 10^{14}\,$cm in \citealt{Slavin2020}), or clump disruption and remnant evolution were not conducted in the same simulation (\citealt{Micelotta2016}). Taking up the last point, we carry out a multi-step approach in our study, in which first the full remnant evolution is modelled in order to determine ejecta and shock properties for different remnant ages, and then a single clump is hit and destroyed by the reverse shock in a so-called cloud-crushing simulation at high resolution.

We aim to model the dust destruction in an oxygen-rich SNR like Cas~A. This dusty SNR has been studied extensively (e.g. \citealt{Dwek1987, Gotthelf2001, Rho2008, Barlow2010,  Milisavljevic2013,Arendt2014, DeLooze2017}) and  provides a unique laboratory to investigate the destruction of dust by a reverse shock. The light from the exploding star likely reached Earth around $1670-1680$, giving the SNR an age of \mbox{$340-350\,$yr} (\citealt{Thorstensen2001}).

In this paper, the time-dependent dust destruction by the reverse shock in a SNR is studied. We conduct a 1D MHD simulation of the expanding ejecta and derive the gas conditions at the position of the reverse shock (Section~\ref{Prag}). These conditions are used to study the evolving environment of an ejecta clump that is impacted by the reverse shock - the MHD simulations of this cloud-crushing problem setup are presented in Section~\ref{Berlin}. In Section~\ref{Amsterdam} we present the dust processing in the shocked and disrupted clumps and derive dust destruction rates and dust masses. We discuss and summarize our findings in Section~\ref{London} and \ref{Dublin}.


\section{SNR expansion}
\label{Prag}
The first step in order to study the impact of the shock at different points during the evolution of the SNR is to determine the exact state of the SN shock front and the ejecta material at each point in time. To obtain this evolution we ran spherically symmetric 1D simulations with \textsc{AREPO} \citep{Volker+2010-AREPO}. The simulation follows the setup of \citet{Truelove1999} where the box is divided into three regions: the ejecta core, an envelope region and the surrounding CSM. The solution for the evolution in the case of no magnetic field and without radiative cooling can be analytically found from three initial conditions, namely the energy of the explosion, $E_\text{SN}$, the ejecta mass, $M_\text{ej}$, and the mass density of the CSM, $\rho_{CSM}$, which the SN material is expanding into. In this model the ejecta core is assumed to have a uniform density distribution, while the envelope density is defined according to a power-law of index $n$, which also determines the structure function $f\left(\frac{v}{v_\text{ej}}\right)$:

\begin{equation}
f\left(\frac{v}{v_\text{ej}}\right) = \begin{cases}
        f_0,                                               &0\le \frac{v}{v_\text{ej}} \le \frac{v_\text{core}}{v_\text{ej}}, \\
        f_\text{n} \left[\frac{v}{v_\text{ej}}\right]^{-n},& \frac{v_\text{core}}{v_\text{ej}}\le  \frac{v}{v_\text{ej}} \le  1,
    \end{cases}
\end{equation}
where $v\sim r/t$ is the expansion velocity at time $t$ and distance $r<v_\text{ej}\,t$,  $v_\text{ej}$ is the maximum initial velocity of the ejecta, $v_\text{core}$ is the velocity of the ejecta at the boundary between core and envelope, and $f_0$ and $f_\text{n}$ are constants. The index $n$ affects the posterior evolution of the SNR such as how long it will take the reverse shock to revert its direction, and the evolution of the velocity of the reverse shock. The value adopted for this index is commonly in the range $5\leq n<10$ (\citealt{Truelove1999,Chevalier2003,Micelotta2016, Bocchio2016}). Here, we use $n=9$ in accordance with \cite{Micelotta2016} and \cite{Laming2003} as this value is known to reproduce well the observables of Cas A by assuming plausible values for the ejecta energy and mass,  $E_\text{SN} = 2.2\times 10 ^{51}\,$ergs and $M_\text{ej} = 2.2\,\text{M}_\odot$, respectively. The mass of $2.2\,\text{M}_\odot$ includes all the gas in the ejecta, with oxygen being the dominant species in Cas~A (\citealt{Willingale2002, Docenko2010}). The density of the CSM is set to $\rho_\text{CSM} = 3.46\times10^{-24}\,\text{g/cm}^3$ which is equivalent to $2.07$ atoms of hydrogen per $\text{cm}^3$, consistent with observations (\citealt{Willingale2003}). Furthermore, these values are the same as used by  \cite{Micelotta2016} which makes it easier to compare to this study.


The mass density of the three defined regions (core, envelope and CSM) can then be expressed as follows:
\begin{equation}
    \rho(r,t) =     \begin{cases}
     \rho_\text{core} = f_0\frac{M_\text{ej}}{v^3_\text{ej}}t^{-3}, & r\le R_\text{core}, \\
     \rho_\text{env} = \frac{M_\text{ej}}{v^3_\text{ej}}f_\text{n}\left[\frac{v}{v_\text{ej}}\right]^{-n} t^{-3}, & R_\text{core}\le r\le R_\text{env}, \\
     \rho_\text{CSM}, &  r\ge R_\text{env}.
    \end{cases} 
\end{equation}
We note that $v_\text{ej}$ can be determined from the energy and mass of the SN explosion since
\begin{equation}
        E_\text{SN}  =  \frac{1}{2}M_\text{ej}v^2_\text{ej}\int_0^1 4\pi \left[\frac{v}{v_\text{ej}}\right]^4 f\left(\frac{v}{v_\text{ej}}\right)  \text{d}\left(\frac{v}{v_\text{ej}}\right),
\end{equation}
and therefore does not add a fourth parameter to the description.



Our simulations differ from the analytical solution (which can be obtained from a given choice of $n$, $E_\text{SN}$, $M_\text{ej}$ and $\rho_\text{CSM}$) in two ways: firstly because we consider the evolution of the magnetic field and secondly because we assume that the SN ejecta cools as an oxygen-only gas since this is the most abundant element in the ejecta of Cas~A \citep{Docenko2010}. For the latter we have incorporated
an oxygen-rich cooling curve  (\citealt{Kirchschlager2019b}, Fig. 3 therein) which combines the cooling curves of \cite{Sutherland1995} for temperatures below $\unit[10^4]{K}$ and the cooling derived by CHIANTI\footnote{https://www.chiantidatabase.org/}
(\citealt{DelZanna2015}) for the temperature range $\unit[10^4-10^9]{K}$. The cooling function derived by CHIANTI is calculated assuming collisional ionisation equilibrium at each gas temperature. We chose a starting value for the magnetic field strength so that the unshocked ejecta at 300$\,$yr has a magnetic field around $\unit[{\sim}1]{\mu G}$ perpendicular to the shock direction, which is amplified to several 100 to a few $1000\,\mu$G in the post-shock gas, consistent with observations of shocked ejecta regions (e.g.~\citealt{Vink2003, Domcek2021}).

      \begin{figure}
 \resizebox{\hsize}{!}{ 
  \includegraphics[trim=1.7cm 1.55cm 2.3cm 1.9cm, clip=true,  page=1]{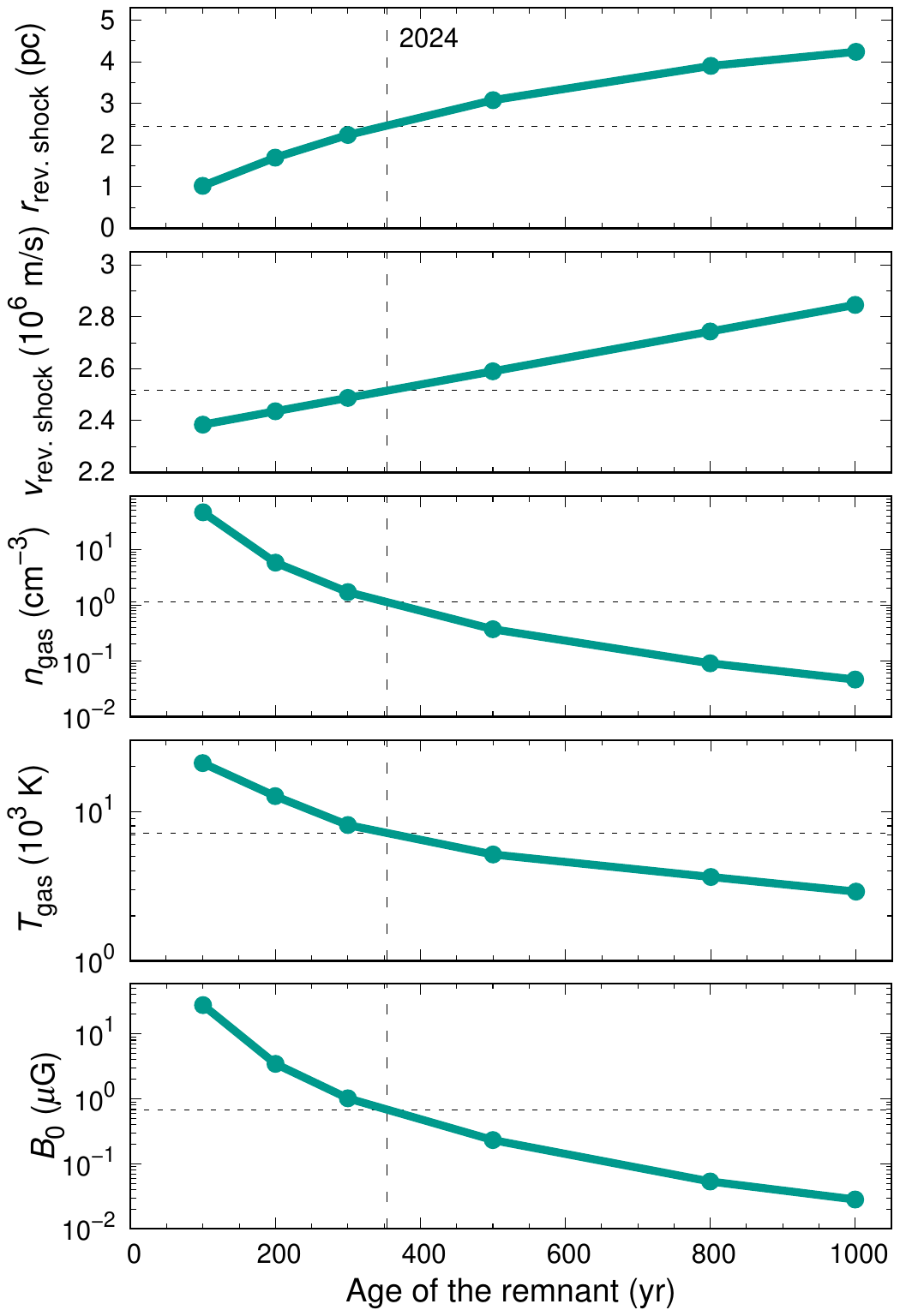}
  }
 \resizebox{\hsize}{!}{ 
  \includegraphics[trim=1.7cm 0.2cm 2.3cm 21.3cm, clip=true,  page=2]{Pics/Evolution_paper.pdf}
  }  
  \caption{\label{Arepo_result} Temporal evolution of the SN explosion in the first $1000\,$yr. First row: Distance $r_\text{rev. shock}$ of the reverse shock in the rest frame of the explosion centre. Second row: Velocity $v_\text{rev. shock}$ of the reverse shock in the frame of the moving ejecta. Third to Fifth row: Pre-shock gas number density $n_\text{gas}$, gas temperature $T_\text{gas}$, and magnetic field strength $B_0$ at the position of the reverse shock. Sixth row: Shocked ejecta mass $M_\text{sh. ej.}$. The vertical line indicates the year 2024; the horizontal line represents the conditions in 2024.}
  \end{figure}

In Fig.~\ref{Arepo_result} the results of the simulation of the SN expansion are shown. The position and velocity of the reverse shock, the shocked ejecta mass as well the gas density, gas temperature and magnetic field strength at the position of the reverse shock are evaluated for six remnant ages: 100, 200, 300, 500, 800, and $1000\,$yr. We can see that the reverse shock is still expanding outwards and the relative velocity between expanding ejecta and reverse shock is still increasing. This means that the collision velocity of gas parcels in the ejecta with the reverse shock increases from ${\sim}2400\,$km/s at $t=100\,$yr to ${\sim}2800\,$km/s at $t=1000\,$yr. For comparison, the reverse shock velocity in the rest frame of the explosion centre decreases continuously over time (see Fig.~\ref{Revshockvelocity} in Appendix~\ref{app4}). In the unshocked ejecta, the gas density and the magnetic field strength drop continuously by three orders of magnitude and the gas temperature by one order of magnitude. On the other hand, the mass of the shocked ejecta increases continuously up to ${\sim}2\,\text{M}_\odot$ within $1000\,$yr. The unshocked ejecta mass at an age of ${\sim}350\,$yr  in the model corresponds well with the estimated $0.5\,\text{M}_\odot$ from observations  (\citealt{Hwang2012, Laming2020}). Since the gas density drops by three orders of magnitude while the shock velocity increases in the same time only by ${\sim}15\,\%$, the energy density of the shock drops by a factor of ${\sim}0.0013$ which has a significant impact on the dust destruction efficiency. We will see in Section~\ref{Amsterdam} that it is sufficient to consider only the first $1000\,$yr of the expansion of Cas~A to properly determine dust destruction in the ejecta. 

\begin{figure}
   \resizebox{\hsize}{!}{
\includegraphics[trim=2.55cm 15.05cm 2.3cm 2.0cm, clip=true,  page=1]{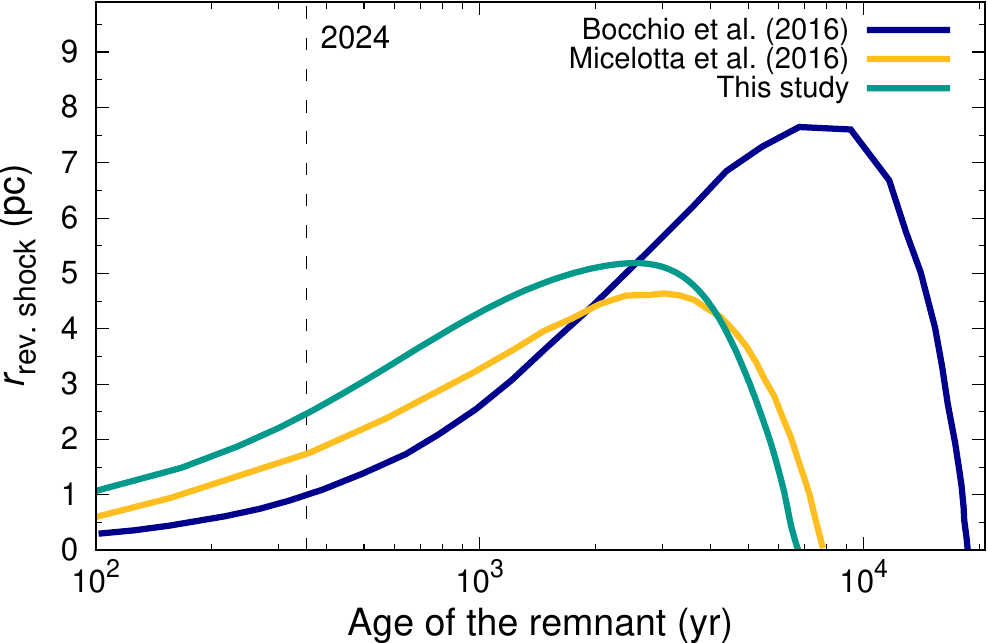}}\vspace*{-0.15cm} 
  \caption{\label{Revshock}Reverse shock position relative to the SN explosion centre from the early evolution until the shock reaches again the explosion centre. The vertical line indicates the year 2024.}
  \end{figure}

In order to compare our SN expansion model with the studies of \cite{Micelotta2016} and \cite{Bocchio2016}, we calculated the reverse shock position for several thousands of years after the SN explosion (Fig.~\ref{Revshock}). In the rest frame of the explosion centre, the shock starts reversing its motion after ${\sim}2450\,$yr and, if the shock could travel unimpeded back to the center of the explosion, this would happen after ${\sim}7600\,$yr. Differences in the shock evolution compared to the studies of \cite{Micelotta2016} and \cite{Bocchio2016} have a significant impact on e.g. the gas density at the reverse shock position and thus on the dust survival rate.

\section{Reverse shock hitting the ejecta clump}
\label{Berlin}
In order to study the destruction of dust in the clumpy ejecta of the SNR, we have at first to introduce the clumps in our model, which was not feasible in the 1D SN explosion model in Section~\ref{Prag}.

The SN ejecta is made up of a relatively smooth ambient (interclump) medium of gas that is intercepted by dense gas clumps. Dust is assumed to have formed exclusively in these clumps as they are cold and dense enough for dust condensation. During the evolution of the SNR, clumps will be hit by the reverse shock.  Different clumps in the ejecta are subject to the reverse shock at different times: the closer the clump to the center, the later it is impacted by the reverse shock. Before the impact, we assume the dense ejecta clumps comove with the interclump medium.

Instead of modelling the full ejecta with a multitude of clumps, we can focus on a section of the remnant in which only one clump is impacted by a planar shock wave. This setup, called the cloud-crushing problem (\citealt{Woodward1976}), allows one to model at higher spatial resolution and was already applied by \cite{Silvia2010, Silvia2012} and \cite{Kirchschlager2019b, Kirchschlager2020, Kirchschlager2023} to study dust destruction in a clumpy SNR. 
 In previous studies on the temporal evolution of dust destruction in SNRs, the question of clump evolution in the pre-shock gas was not properly addressed -- if clumps were considered at all (see Section \ref{London} for an overview of previous studies). In order to introduce clumps to our model, we assume a fixed clump gas temperature and fixed clump mass for each clump and for each time of the SN evolution as well as gas pressure equilibrium between the clumps and interclump medium. As a consequence, the clump gas density decreases and the clump radius increases over time.

The results of the 1D SN explosion model in Section~\ref{Prag} define the gas conditions of the unshocked interclump medium. This is applicable because we assume that the reverse shock velocity is unaffected by the presence of clumps in the ejecta. The gas conditions of the unshocked clumps are defined by the parameters summarized below.
  
  
  In accordance with the six time-steps from the 1D SN explosion model, we run six cloud-crushing simulations. The \textit{initial} conditions of the six cloud-crushing simulations are defined as the following:
  The initial gas density $n_\text{ic}$ and gas temperature $T_\text{ic}$ of the interclump medium as well as the shock velocity $v_\text{sh}$ and the magnetic field strength $B_0$ are taken from the SN expansion simulation (Section~\ref{Prag}) for each remnant age. The pre-shock magnetic field orientation in the interclump medium and in the clump is perpendicular to the shock direction. A parallel alignment would have a significantly lower effect on the dust destruction rate (\citealt{Kirchschlager2023}). We assume that the embedded clump is spherical. Clump gas temperature and density are not defined by the SN expansion simulation and have to be inferred differently. \cite{Micelotta2016} found that the temperature in the shocked ejecta drops within ${\sim}0.5\,$yr from $3\times10^6\,$K to $300\,$K due to cooling. Since this timescale is much shorter than the remnant evolution time, it is reasonable to assume that also  the temperature of the unshocked clumps is not higher than a few $100\,$K. From an observer perspective, \cite{Raymond2018} estimated a temperature of $100\,$K from [Si I]/[Si II] line ratios for the coolest part of the unshocked ejecta in Cas~A. This value is in line with previous modelling papers that studied the clump-crushing setup in SNRs (\cite{Silvia2010, Silvia2012, Kirchschlager2019b, Kirchschlager2020, Kirchschlager2023}. Therefore, the clump gas temperature in our model is fixed to \mbox{$T_\text{cl}=\unit[100]{K}$}.
  Following pressure equilibrium between clump and interclump medium, the homogeneous clump gas density then amounts to $n_\text{cl} = (T_\text{ic}/T_\text{cl}) n_\text{ic}$.\footnote{We do not assume homologous clump expansion as this would imply higher clump gas temperatures at young ages, which is not consistent with the effective cooling in Cas~A. We note that the exact description of the behavior of an expanding, homogeneous gas clump within an expanding ejecta under the influence of cooling has to be modelled by its own for different clump sizes, gas densities and gas temperatures, which is beyond our study.} The size of the clump is a function of time $t$. Clumps in observations of Cas~A show `today' radii in the range of $\unit[(0.5-2.5) \times 10^{16}]{cm}$ (\citealt{Fesen2011}). Adopting an age of $\unit[{\sim}340-350]{yr}$ for  Cas~A (\citealt{Fesen2006}), we set the radius of the spherical clump  at $t=\unit[300]{yr}$ to $R_\text{cl} = \unit[10^{16}]{cm}$. The clump radius at time $t$ is then $R_\text{cl}(t) =  (n_\text{cl}(t=\unit[300]{yr})/n_\text{cl}(t))^{\nicefrac{1}{3}} \times \unit[10^{16}]{cm}$.   The mean molecular weight of the pre-shock gas is set to $\mu = 16.0$, corresponding to a pure oxygen gas as in Cas~A, and the adiabatic exponent is $\gamma_\text{hydro} = 5/3$.
  
\begin{figure}
 \resizebox{\hsize}{!}{ 
  \includegraphics[trim=1.7cm 9.85cm 2.3cm 1.9cm, clip=true,  page=3]{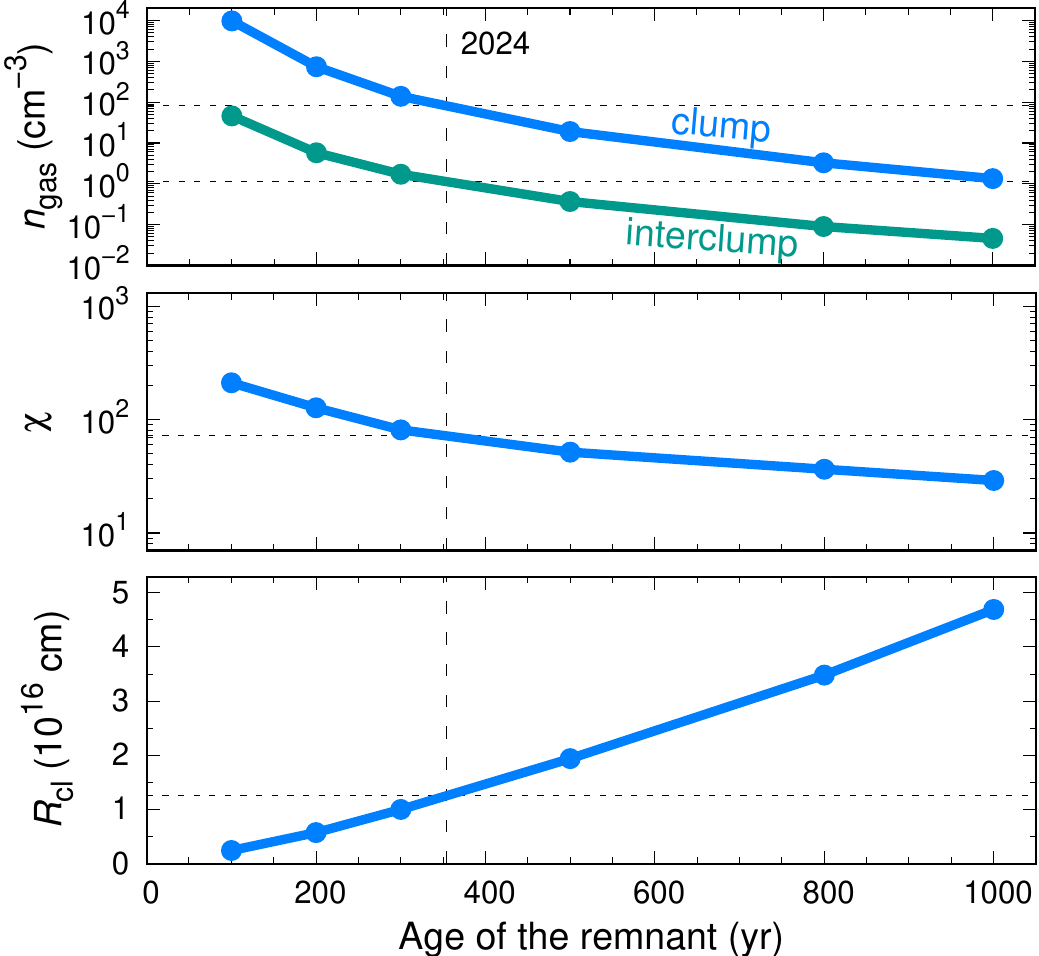}
  }
  \caption{\label{AB_initial} Temporal evolution of the pre-shock clump properties (= initial conditions of the cloud-crushing setup): Gas number density $n_\text{gas}$ of the interclump medium (green) and clump (blue; first row), density contrast $\chi$ between clump and interclump medium (second row), and clump radius $R_\text{cl}$. Vertical and horizontal lines as in Fig.~\ref{Arepo_result}.}
  \end{figure}

  The gas density of clump and interclump medium, the density contrast $\chi=n_\text{cl}/n_\text{ic}$, and the clump radius for each time $t$ after the SN explosion are shown in Fig.~\ref{AB_initial}. We see that the clump radius continuously increases from $\unit[0.24 \times 10^{16}]{cm}$ at $t= \unit[100]{yr}$ to $\unit[4.69 \times 10^{16}]{cm}$ at $t= \unit[1000]{yr}$, while the density contrast decreases in the same time interval from $211$ to $29$. We note that an extrapolation to the initial years of the explosion (${\ll}100\,$yr) is not suitable, as this would imply extremely high density contrasts and extremely low clump filling factors (see Figs.~\ref{AB_initial}). We will see in Section~\ref{Amsterdam} that the reverse shock arises at an age of ${\sim}63\,$yr, which is in the range observed in several SNRs. We can ignore any clump and remnant evolution before the reverse shock occurs for the first time.
  
Based on these initial (pre-shock) conditions, we use the MHD code \textsc{\mbox{AstroBEAR}}\footnote{\href{https://www.pas.rochester.edu/astrobear/}{https://www.pas.rochester.edu/astrobear/}} (\citealt{Cunningham2009, Carroll-Nellenback2013}) to simulate the dynamical evolution of a reverse shock impacting a clump of ejecta material in the SNR and to calculate the post-shock conditions. In agreement to the SN expansion simulations in Section~\ref{Prag}, the cloud-crushing simulations consider cooling of the oxygen-rich gas (\citealt{Kirchschlager2019b}, Fig. 3 therein).  Each cloud-crushing simulation is executed for a time  of $3\,\tau_\text{cc}$ after the first contact of the shock with the clump, where the cloud-crushing time $\tau_\text{cc} = \chi^{0.5} R_\text{cl}/v_\text{sh}$ (\citealt{Klein1994}) is the characteristic timescale for the clump to be crushed  and $3\,\tau_\text{cc}$ is a standard value to investigate the post-shock structures. The total simulation time of the cloud-crushing setup amounts to $t_\text{sim} = \unit[14.4]{yr}$ at $\unit[100]{yr}$ and rises up to $\unit[89.7]{yr}$ at $\unit[1000]{yr}$ after the SN explosion. We consider 2D MHD simulations due to the large computational effort for highly resolved 3D post-processing simulations (following in Section~\ref{Amsterdam}).  Based on previous studies (see \citealt{Kirchschlager2023}), the length and width of the domain are set to $l_\text{box} = 15\, R_\text{cl}$ and  $w_\text{box} = 6\,R_\text{cl}$  which ensures that the material of the disrupted clump does not flow out of the domain at the back end during the simulation time $t_\text{sim}$. The computational domain of all six cloud-crushing simulations consists of $1500 \times 600$ cells such that the physical resolution is between $\Delta_\text{cell} = \unit[0.24\times10^{14}]{cm}$ $({\sim}\unit[1.6]{au})$ per cell at $\unit[100]{yr}$ and $\unit[4.69\times10^{14}]{cm}$  $({\sim}\unit[31.4]{au)}$ at $\unit[1000]{yr}$ after the SN explosion. On all sides of the domain, outflow boundary conditions are used  with the exception of the left boundary which used an inflow condition for injecting a continuous post-shock wind into the domain.

\begin{figure*}
   \resizebox{0.83\hsize}{!}{
\includegraphics[trim=2.7cm 7.03cm 5.4cm 0.8cm, clip=true,  page=1]{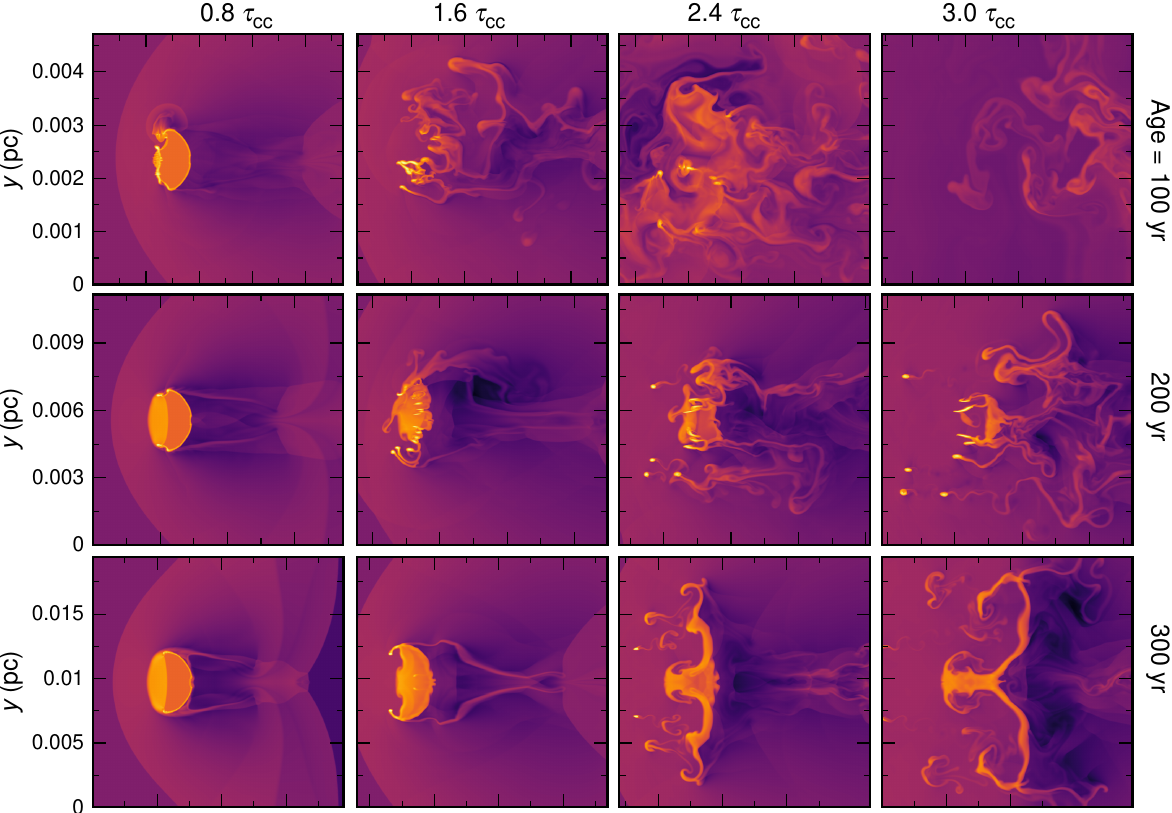}}\\
 \resizebox{0.83\hsize}{!}{\includegraphics[trim=2.6cm 6.3cm 5.4cm 1.35cm, clip=true,  page=2]{Pics/Overview_maps.pdf}}\vspace*{-0.15cm} 
  \caption{\label{Clumpmaps}Gas number density of clumps at different remnant ages (rows). The maps show the clump disruption at $0.8\,\tau_\text{cc}$, $1.6\,\tau_\text{cc}$, $2.4\,\tau_\text{cc}$, and $3.0\,\tau_\text{cc}$ (from left to right). The color scale is fixed for each row.}
  \end{figure*}

 The gas density maps of the shocked and disrupted gas clump are shown for different remnant ages and snapshots in Fig.~\ref{Clumpmaps}. The disruption of clumps and clouds by shock waves has been studied and analysed before. For discussions, we refer to \cite{Bedogni1990, Stone1992, Klein1994, Orlando2005, Orlando2006, Agertz2007} and, in particular for SNR clumps impacted by the reverse shock, to \cite{Silvia2010, Silvia2012} and \cite{Kirchschlager2019b,Kirchschlager2020, Kirchschlager2023}. In the following Section we will use the spatially and temporally resolved data of the shocked and disrupted gas clump and post-process them in order to investigate the dust evolution.

   
\section{Dust destruction in the shocked clump}
\label{Amsterdam}

We assume a homogeneous dust distribution in the unshocked clump while the unshocked interclump medium is dust-free. In order to investigate the grain dynamics and the dust destruction in the post-shock gas, we use the dust processing code \textsc{\mbox{Paperboats}} (\citealt{Kirchschlager2019b,Kirchschlager2023}). \textsc{\mbox{Paperboats}} is a parallelized 3D post-processing code that computes the evolution of the dust on the basis of the temporally and spatially resolved gas
density, velocity, temperature, and magnetic field output of a MHD simulation, as well as on the initial dust conditions. We give here a summary of the considered processes.

The dust is accelerated by the gas taking into account both collisional and plasma drag (\citealt{Baines1965, Draine1979}). Charged grains are also accelerated by the magnetic field due to the Lorentz force (e.g. \citealt{Fry2020}). Destruction processes consider fragmentation and vaporization in grain-grain collisions (e.g. \citealt{Jones1996, Hirashita2009}) as well as thermal and non-thermal (kinematic) sputtering (e.g. \citealt{Barlow1978, Tielens1994}). Grain growth processes include the accretion of dusty gas\footnote{Dusty gas is atomic and composed of destroyed dust grain material, produced by the (partial) destruction of grains in previous destruction events. At the start of the simulation there is no dusty gas.} onto the surfaces of the grains, ion-trapping of dusty gas (\citealt{Kirchschlager2020})  as well as the coagulation (sticking; e.g. \citealt{Chokshi1993}) of dust grains in low-velocity grain-grain collisions. The latter is negligible for the dust processing as the present grain velocities are mostly too high. We ignore the accretion and ion-trapping of nebular oxygen gas as this can cause unrealistically large dust grain growth due to the nature of the post-processing (see  \citealt{Kirchschlager2020}). Other effects considered are size-dependent sputtering (\citealt{SerraDiazCano2008}), partial grain vaporization (\citealt{Kirchschlager2023}),  Coulomb interaction between charged grains and the ionised gas, as well as grain charging due to impinging electrons and ions, secondary electron emission, transmitted electrons, and field emission (\citealt{Fry2020}). Thermal and non-thermal sputtering are combined by a skewed Maxwellian velocity distribution (\citealt{Bocchio2014}). 
Due to the nature of the post-processing any feedback from the dust on the gas, like momentum transfer, dust cooling or heating, has to be ignored. 

In this study, the MHD simulations are in 2D, however, it is crucial to consider grain-grain collisions in 3D as this will affect the grain cross-sections and collision probabilities. The 2D MHD simulations are extended to 3D assuming a single cell in the \mbox{z-direction}, spanning a volume of $1500\times600\times1$ cells (see Section~\ref{Berlin}), as well as assuming no gas velocity in the \mbox{z-direction}. The clump shape for the post-processing  is thus a flat 3D circular disc. We have seen in \cite{Kirchschlager2020} at a lower resolution that the 2D and 3D gas distributions of the cloud-crushing problem show only moderate differences and yield similar dust survival fractions.

The dust material is silicate and the material parameters required for the dust processing are given in Table 2 of \cite{Kirchschlager2019b}. The grain sizes can cover a range from $0.6\,$nm to $3\,\mu$m and are discretized in 32 log-spaced size bins. Furthermore, we consider two additional collector bins at the lower and upper end of the size range to track the material that crosses the size range limits. The material in the lower collector bin represents the destroyed dust which is equivalent to the dusty gas. Dust material is called destroyed if it is vaporized or sputtered to the gas phase or if the radius of the grain is below $0.6\,$nm, otherwise the dust survives.

\begin{figure}
 \resizebox{\hsize}{!}{ 
  \includegraphics[trim=2.1cm 15.3cm 2.3cm 2.0cm, clip=true,  page=1]{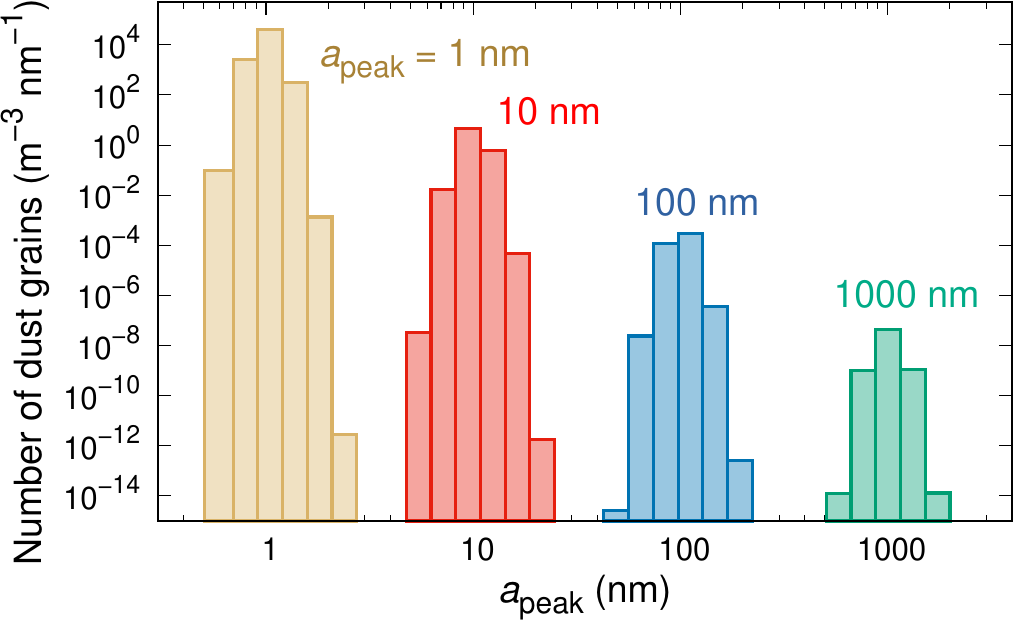}
  }
  \caption{\label{init_gsd} Initial grain size distributions in the pre-shock gas at an remnant age of $100\,$yr for peak radii $a_\text{peak}=1,10,100$ and $1000\,$nm and distribution width $\sigma=0.1$.}
  \end{figure}

Initially, the spherical grain sizes follow a log-normal distribution (e.g. \citealt{Nozawa2003}) that is described by their radius $a_\text{peak}$ at which the distribution has its maximum and the dimensionless quantity $\sigma$ that defines the width of the distribution. For a progenitor mass of $15-25\,\text{M}_\odot$, dust formation theories predict silicate grains to have radii of ${\sim}1-100\,$nm (\citealt{Todini2001, Nozawa2003, Bianchi2007,  Marassi2015, Sarangi2015, Biscaro2016, Bocchio2016}), while the grain radii derived from modelling observational data are of the order of $0.1\,\mu$m up to a few micrometres (\citealt{Stritzinger2012, Gall2014, Fox2015, Owen2015, Wesson2015,  Bevan2017, Priestley2020}). Therefore, grain size ranges of four orders of magnitude should be considered when studying dust in the ejecta of supernova remnants (SNRs). We vary $a_\text{peak}$ between $1$, 10, 100, and $1000\,$nm and fix  $\sigma=0.1$ (Fig.~\ref{init_gsd}). At the start of the simulation, the dust is exclusively in the clump, homogeneously distributed, and at rest compared to the gas. In order to represent Cas~A like conditions, the gas-to-dust mass ratio in the clump is set to $\Delta_\text{gd} = 10$ (\citealt{Priestley2019a}). The dust mass of a single clump then amounts to $7.78\times10^{-7}\,\text{M}_\odot$ in the model. The total dust mass in Cas~A has been derived by different
observational strategies to be between ${\sim}0.1$ and $1\,\text{M}_\odot$ (\citealt{Dunne2009, Barlow2010, Bevan2017, DeLooze2017, Priestley2022b}). Consequently,  ${\sim}10^5-10^6$ clumps are required in the ejecta remnant.

\begin{figure}
 \resizebox{\hsize}{!}{ 
  \includegraphics[trim=1.7cm 14.7cm 2.3cm 1.83cm, clip=true,  page=4]{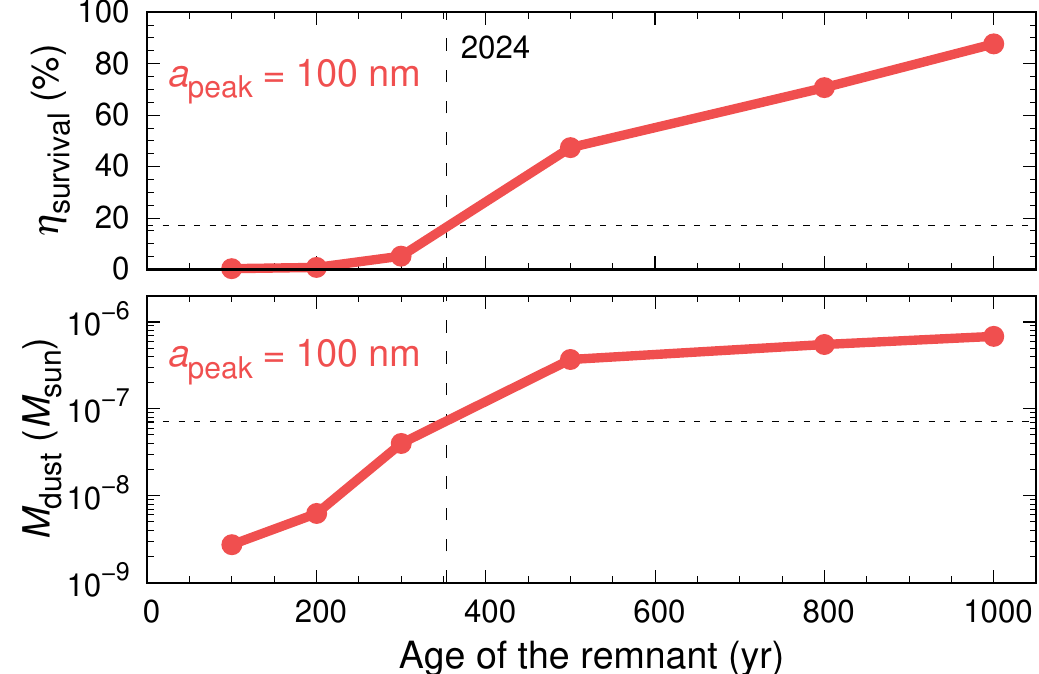}
  }
  \caption{\label{Pap_results} Temporal evolution of the dust survival fraction $\eta_\text{survival}$ (top) and the total surviving dust mass $M_\text{dust}$ (bottom), both for a single clump meeting the reverse shock at different remnant ages. The initial grain sizes follow a log-normal distribution with peak radius $a_\text{peak}=\unit[100]{nm}$. Vertical and horizontal lines as in Fig.~\ref{Arepo_result}.}
  \end{figure}

We run the post-processing simulations of the cloud-crushing simulations (Section~\ref{Berlin}) and calculate the dust density maps, grain size distributions and dust destruction rates for each grid cell and each time-step. The dust mass survival fractions $\eta_\text{survival}$ and the total surviving dust mass of a single clump for the grain size distribution with $a_\text{peak}=100\,$nm are shown in Fig.~\ref{Pap_results}. It can be seen that the survival fraction is continuously increasing as a function of the remnant age when the clump encounters the reverse shock. Less than 1$\,$per~cent of the initial dust mass is surviving in the first $200\,$yr, while more than 87$\,$per~cent survives the reverse shock when it hits the clump at a SNR age of $1000\,$yr. The surviving dust mass\footnote{We note that the dust masses are calculated for a 3D spherical clump by extrapolating from the surviving dust masses of the 3D circular disc.} in the single clump is proportional to the survival fraction and thus increases continuously from $2\times10^{-9}$ to $6.82\times10^{-7}\,\text{M}_\odot$ from 100 to $1000\,$yr (compared to $7.78\times10^{-7}\,\text{M}_\odot$ as the original dust mass per clump). In 2024, 17$\,$per~cent of the initial dust mass would survive the impact, which corresponds to $1.32\times10^{-7}\,\text{M}_\odot$. The reduced dust destruction over time is consistent with the decreasing energy of the reverse shock, mainly forced by the reduced gas density (Fig.~\ref{Arepo_result}). The reduced impact energy causes lower gas and dust velocities in the post-shock gas, which is crucial in reducing the efficiency of sputtering and grain-grain collisions.

\begin{figure}
 \resizebox{\hsize}{!}{ 
  \includegraphics[trim=2.4cm 15.07cm 2.3cm 1.82cm, clip=true,  page=1]{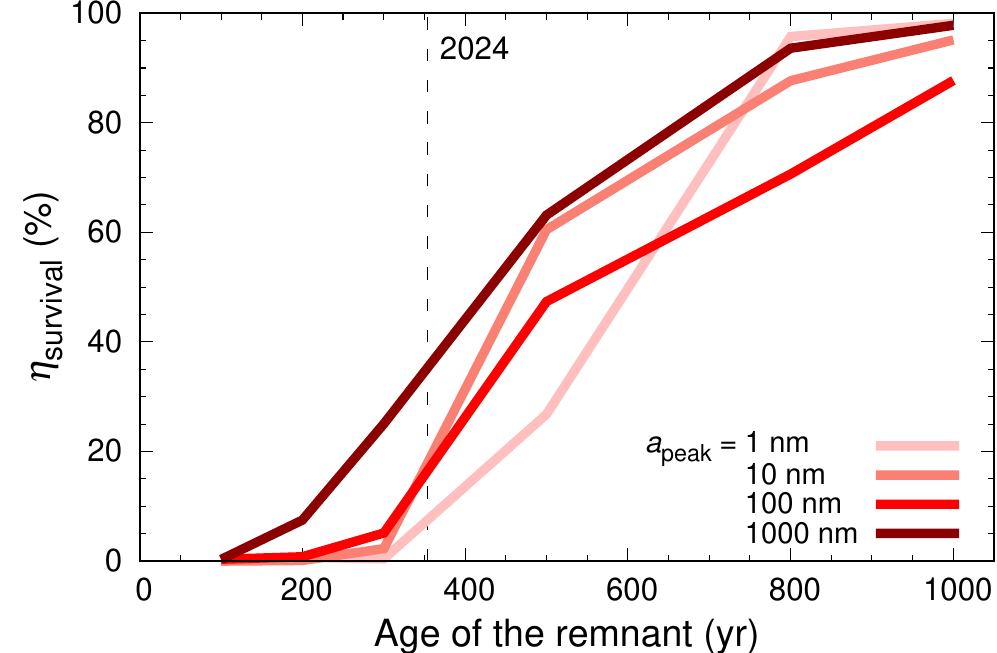}
  }
  \caption{\label{different_grain_sizes} Temporal evolution (in a single clump meeting the reverse shock at different remnant ages) of the dust survival fraction $\eta_\text{survival}$ for peak radii $a_\text{peak}=1,10,100$ and $1000\,$nm. Vertical line as in Fig.~\ref{Arepo_result}.}
  \end{figure}

The survival fractions of different grain sizes (Fig.~\ref{different_grain_sizes}) are at a similar level and show the same qualitative behavior as for \mbox{$a_\text{peak}=100\,$nm}. Size distributions with \mbox{$a_\text{peak}=1\,$nm} are slightly more efficiently destroyed than bigger grains in the first ${\sim}600\,$yr, before they reach a turning point ($800\,$yr) after which they are less efficiently destroyed than large particles. On the other hand, the survival fraction of size distributions with \mbox{$a_\text{peak}=1000\,$nm} exceeds the 1$\,$per~cent level already briefly after $100\,$yr but the slope of the curve is flatter than for small grains. The reason for the observed differences is the influence of the grain size on the dust dynamics (drag and Lorentz force) and on the destruction processes (sputtering and grain-grain collisions). The dependencies are complex; we refer to Fig.~20 in \cite{Kirchschlager2023} for an illustrative overview of the efficiency of sputtering, grain-grain collisions and gas drag for different grain sizes and clump gas densities.

Although we did not run simulations without magnetic field, we can estimate its impact on the dust destruction. \cite{Kirchschlager2023} investigated the dust survival rates for different magnetic field strengths in the clumpy ejecta of Cas~A. They found a reduced dust survival rate when the magnetic field increases from 0 over $1\,\mu$G up to $10\,\mu$G. In our model, magnetic field strengths above $1\,\mu$G occur only at young remnant ages (${\lesssim}\,300\,$yr; Fig.~\ref{Arepo_result}). Therefore, we can expect that the influence of magnetic fields is negligible for remnant ages above $300\,$yr. At younger ages, magnetic fields might have a slightly larger impact. However, the dominant effect at early stages is the high gas density and thus the high shock energy which destroy almost the entire dust mass in the clumps encountering the reverse shock. 

\begin{figure}
 \resizebox{\hsize}{!}{ 
  \includegraphics[trim=1.3cm 2.0cm 0.7cm 4.1cm, clip=true,  page=1]{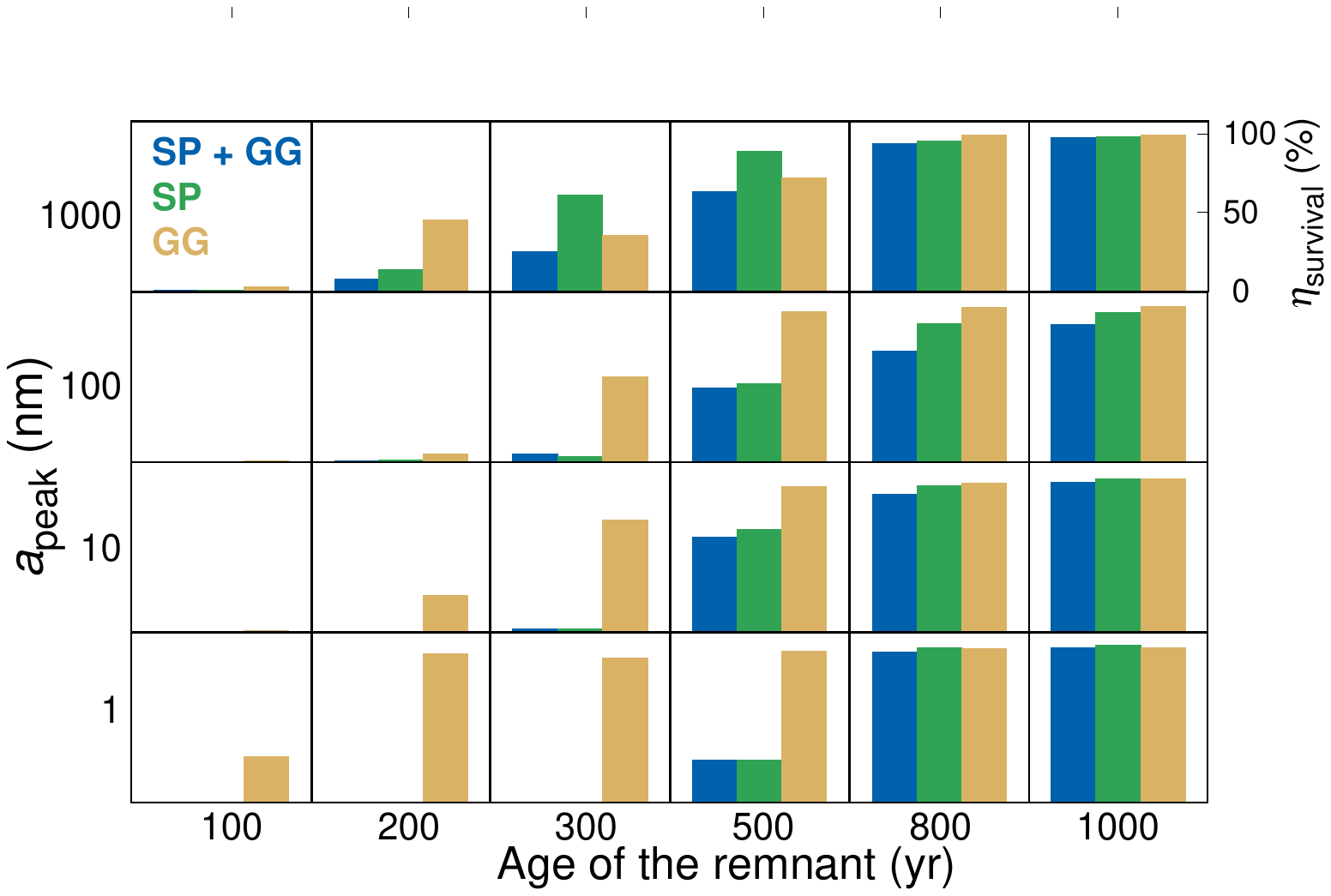}
  }
  \caption{\label{pic_sp_gg} Dust survival fraction $\eta_\text{survival}$ if the destruction processes are considered individually: Sputtering (SP, green), grain-grain collisions (GG, yellow), and the combined effects of sputtering and grain-grain collisions (SP + GG, blue). The results are shown as a function of remnant age (horizontal axis) and peak radius $a_\text{peak}$ of the initial grain size distribution (vertical).}
  \end{figure}

We ran the post-processing simulations again for the same setup but taking into account only sputtering or only grain-grain collisions as destructive processes. Fig.~\ref{pic_sp_gg} shows a bar chart diagram for the survival fraction when the destruction processes are considered individually or combined. The survival fraction increases continuously with increasing remnant age when sputtering and grain-grain collisions are combined, but also for the case of pure sputtering. Sputtering is the dominant factor for most grain sizes and remnant ages, however, grain-grain collisions additionally reduce the survival fraction. This effect becomes most effective for 100 and $1000\,$nm grains. In particular, for $a_\text{peak}=1000\,$nm at 300 and $500\,$yr, grain-grain collisions have an even stronger effect than  sputtering.

       \begin{figure}
   \resizebox{\hsize}{!}{
  \includegraphics[trim=1.7cm 15.07cm 2.3cm 1.68cm, clip=true,  page=1]{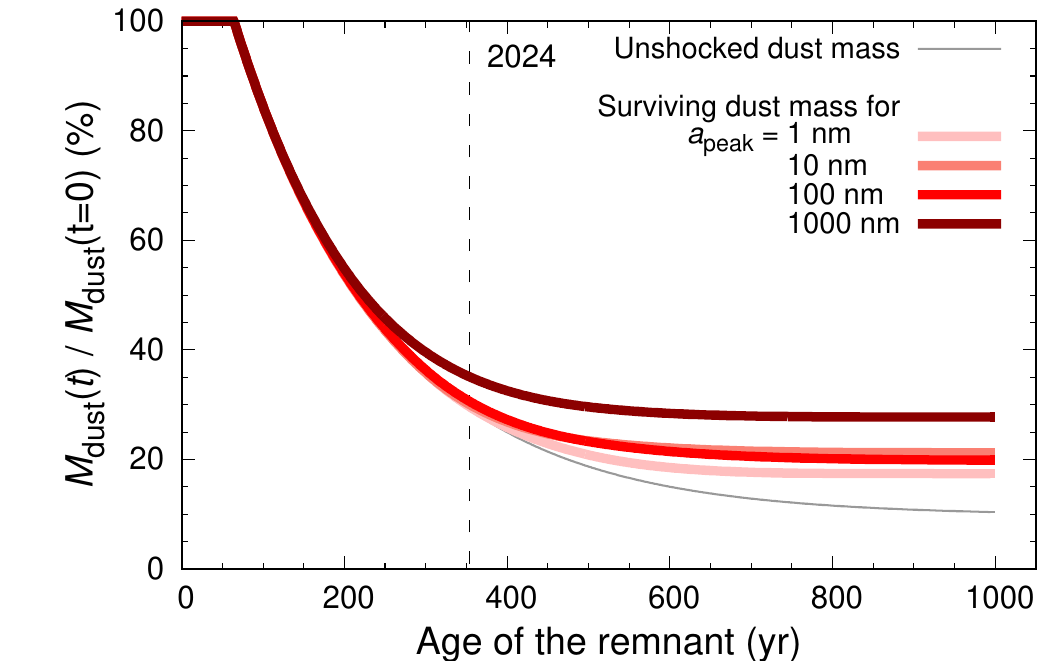}}
   \resizebox{\hsize}{!}{
  \includegraphics[trim=1.4cm 15.07cm 2.3cm 1.68cm, clip=true,  page=2]{Pics/Clump_frequency_paper.pdf}}  
  \caption{\label{Clump_frequency} Survival and destruction integrated over the entire SNR. Top: Dust mass as a function of grain size (coloured lines) and remnant age, normalized by the initial dust mass. The grey solid line represents the dust mass that has not yet passed the reverse shock. Bottom: Fractional dust destruction rates as a function of remnant age and for different grain sizes. Vertical lines as in Fig.~\ref{Arepo_result}.}
  \end{figure}

We are interested in the amount of dust that is processed not only in a single clump but over the entire SNR. For this purpose we assume that the number density of ejecta clumps is proportional to the gas density at each time and at each position in the unshocked remnant. As a consequence, gas and dust mass in the unshocked ejecta scale in the same way and gas masses can be used (instead of the dust masses) to calculate the cumulative dust survival rate in the remnant:
\begin{align}
 \frac{M_\text{dust}(t)}{M_\text{dust}(t=0)} =  \frac{M_\text{unsh. ej.}(t)+ \int_{0}^t \left(\frac{\text{d}M_\text{sh. ej.}(t')}{\text{d}t}\eta_\text{survival}(t')\right) \text{d}t'}{M_\text{ej}}, \label{Krakau}
\end{align}
where $M_\text{dust}(t)$ is the dust mass as a function of remnant age, $M_\text{dust}(t=0)$ is the initial dust mass, $M_\text{ej}=2.2\,\text{M}_\odot$ is the gas mass in the ejecta, $M_\text{unsh. ej.}(t)$ and $M_\text{sh. ej.}(t)$  are the gas masses of the unshocked and shocked ejecta as a function of remnant age, respectively,  and $\eta_\text{survival}(t)$ is the previously derived dust survival rate as a function of remnant age. We fit the gas masses of the shocked ejecta derived from the SNR expansion simulation (Fig.~\ref{Arepo_result}, bottom panel) with the function 
\begin{align}
M_\text{sh. ej.}=1.985\,\text{M}_\odot-2.792\,\text{M}_\odot\,\exp{(-0.005304\,t/\text{yr})}.\label{eq_fitM}
\end{align}
Using $M_\text{unsh. ej.}(t)= M_\text{ej} - M_\text{sh. ej.}(t)$ and eq.~(\ref{eq_fitM}), eq.~(\ref{Krakau}) can be applied to calculate the dust mass for each remnant age. Please note that we assume that the dust is completely formed before the reverse shock impacts the first clumps.

We see in Fig.~\ref{Clump_frequency} (top panel) the amount of dust that has not yet passed the reverse shock (grey solid line) and the dust mass in the remnant as a function of grain size (coloured lines) and remnant age. Following the fit equation for the shocked ejecta mass, the reverse shock starts to process the material at an age of ${\sim}63\,$yr. After that, the amount of dust existing in the remnant is continuously decreasing with age due to the passage of the reverse shock. 50 per~cent of the dust has already passed the reverse shock at the age of 217 yr and 70 per~cent of the dust by the year 2024. Since the dust destruction rate in the clumps is very high at the early stages (Fig.~\ref{different_grain_sizes}), between 65 and 70 per~cent of the total ejecta dust is already destroyed by 2024. This value is in agreement with the Cas~A dust destruction fraction derived from \textit{Herschel} observations (${\sim}70\,$per~cent; \citealt{DeLooze2017}) and slightly lower than the destruction fraction derived from SED modelling ($74-94\,$per~cent; \citealt{Priestley2022b}).

\begin{figure*}
   \resizebox{0.88\hsize}{!}{
  \includegraphics[trim=1.7cm 15.0cm 2.3cm 1.68cm, clip=true,  page=3]{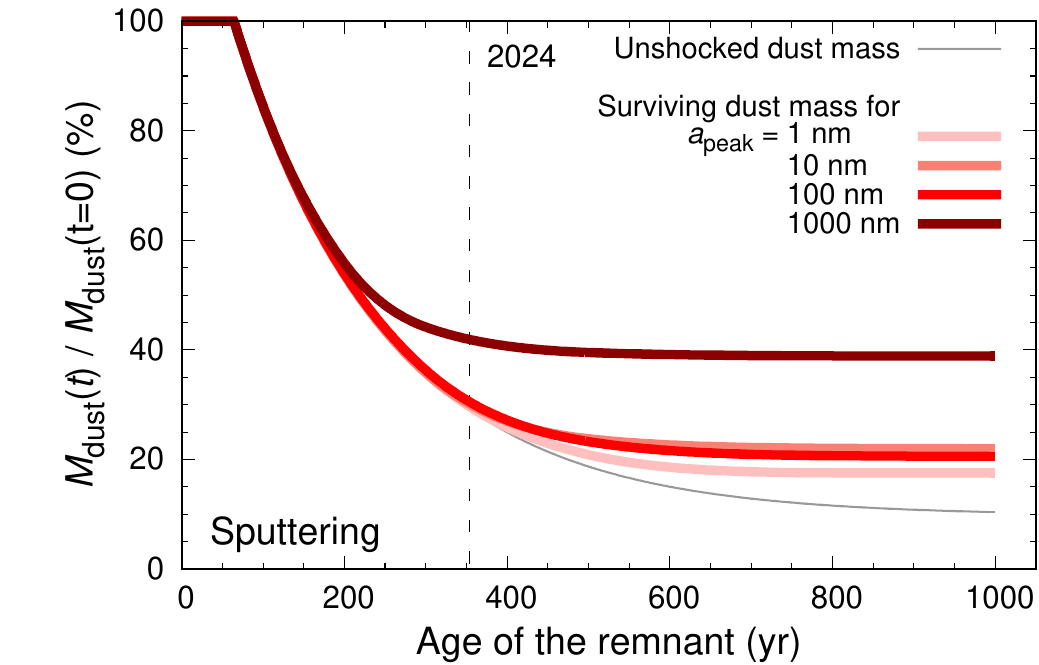}
  \includegraphics[trim=4.6cm 15.0cm 2.3cm 1.68cm, clip=true,  page=4]{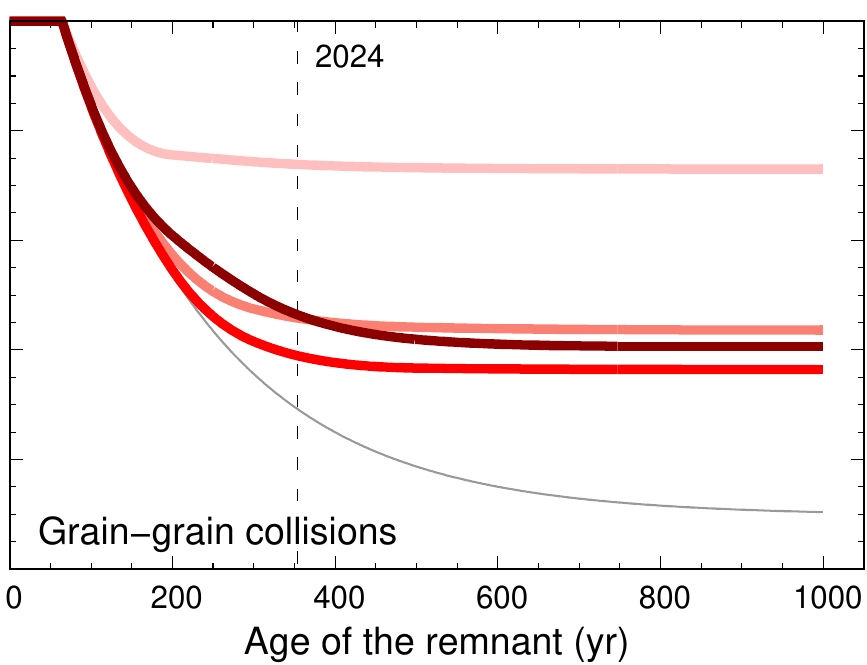}}\vspace*{-0.15cm}   
\caption{\label{Clump_frequency2} Dust destruction integrated over the entire SNR if only sputtering (left) or grain-grain collisions (right) are considered. The grey solid lines represent the dust mass that has not yet passed the reverse shock. The vertical lines indicate the year 2024.}
  \end{figure*}

Since there is nearly no dust surviving in the clumps at the very beginning, there is also no dependency on the grain size. With increasing time the dust survival rates per clump ($\eta_\text{survival}$) increase and thus the dust mass existing in the remnant depends more and more on the grain size.  At ${\sim}600\,$yr the dust masses start to saturate when the survival rates in the clumps strongly increase and at the same time less and less dust is passing the reverse shock. After $1000\,$yr, only $0.2\,\text{M}_\odot$ of gas are still unshocked and thus less than 10 per~cent of the initial dust mass has not yet passed the reverse shock. In a rough trend the surviving dust mass is higher for large grains (17 per~cent for $a_\text{peak}=1\,$nm vs. 28 per~cent for $1\,\mu$m) which is due to the decreasing efficiency of sputtering of large grains. This trend is interrupted by the $100\,$nm grains for which grain-grain collisions have a bigger impact. Furthermore, at later ages, the coupling between gas and dust grains decreases significantly due to the reduced gas densities, and the $100\,$nm grains are more quickly ejected from the shocked clump fragments into the hot interclump medium than smaller grains. This results for the $100\,$nm grains in a slightly lower surviving dust fraction than for the $10\,$nm grains (20 vs. 21 per~cent). The surviving dust masses of the individual destruction processes, sputtering and grain-grain collision, are shown in Fig.~\ref{Clump_frequency2}.


%
%
%
%

The temporal change of dust mass in the SNR (normalized by the original dust mass $M_\text{dust}(t=0)$) is shown in the bottom panel of Fig.~\ref{Clump_frequency}. The destruction rate is continuously decreasing: For a remnant age of $100\,$yr, $5\times10^{-3}M_\text{dust}(t=0)$ is destroyed per year, while for a remnant age of $1000\,$yr only ${\sim}10^{-6}\,M_\text{dust}(t=0)$ is destroyed annually. Because of these low destruction rates at the end of our simulation and the continuing trend, we expect no significant dust destruction after $1000\,$yr and we can ignore the processing at older remnant ages.


\section{Discussion}
\label{London}
\subsection{Previous studies}
To our knowledge, there have been four studies in the past 10 years that have investigated the destruction of freshly produced dust in an expanding SNR considering the evolving conditions of the ejecta gas and of the shocks. We confine the following discussion to studies with a proper remnant evolution and do not take into account those that determined dust destruction rates at different shock velocities.

\cite{Micelotta2016} used the analytical model of \cite{Truelove1999} to describe the evolution of the reverse shock in Cas~A and calculated the dust destruction from the first appearance of the reverse shock until it reaches the SNR center at ${\sim}8000\,$yr (see Fig.~\ref{Revshock}). Similar to our study they find a high amount of dust destruction in the early years which is decreasing over time. Their cumulative dust survival fraction of silicate dust starts to saturate after ${\sim}3000\,$yr and converges against $9\,$per~cent at ${\sim}8000\,$yr. Their survival fractions are lower compared to our findings, however, a comparison to our study is difficult as a few different conditions and processes are considered. On the one hand, \cite{Micelotta2016} did not consider the evolution of the clumps in the pre-shock gas and did not model the clump disruption. Instead, all dust grains are assumed to be exposed to the harsh conditions in the shocked, hot interclump medium: some of them immediately after the first contact of the shock with the clump, but at the latest after $3.5$ cloud-crushing times all particles, which implies a higher destruction rate compared to our study in which a significant part of the grains can still remain in dense clump fragments. On the other hand, \cite{Micelotta2016} ignored grain-grain collisions and gyromotion of charged dust grains around magnetic field lines,  considered kinetic sputtering only for grains as long as they are inside the clumps and thermal sputtering only for dust grains that escaped from the clumps, and took into account the continuing sputtering of grains in the shocked interclump medium beyond a few cloud-crushing times, all of which would otherwise have resulted in an even lower survival rate. Their initial grain-size distribution is MRN-like (\citealt{Mathis1977}). Extrapolation of their results to results of log-normal size distributions should be done with caution as the many small particles in the MRN distribution are completely destroyed by sputtering, while the few large particles have high survival rates.

\cite{Bocchio2016} also used analytical approximations for the evolution of a SNR based on the work by \cite{Truelove1999} to study the dust processing over the full remnant age. 
Apart from magnetic field effects, they considered the same dust destruction processes as in our study and additionally also dust sublimation, which however has a negligible impact on the dust survival rates. Based on their setup, the reverse shock needs ${\sim}20\,$kyr to return to the explosion center (see Fig.~\ref{Revshock}). As a consequence of the slower remnant evolution, their dust destruction seems to be delayed compared to observations and other studies. For Cas~A, for example, they find a dust survival fraction of ${\sim}90\,$per cent at stages as observed `today' (2000 -- 2020) which distinctly exceeds the values derived from \textit{Herschel} observations (\citealt{DeLooze2017}) or SED modelling (\citealt{Priestley2022b}). In the study of \cite{Bocchio2016} the highest dust destruction is between a few hundred and $10^4$ yr and continues until $1\,$Myr in the post-shock gas. Strikingly, the largest dust grains in their model are able to pass the contact discontinuity between reverse and forward shock and even the forward shock, permanently being processed and reduced in size. After $1\,$Myr, only ${\sim}1\,$per cent of the initial dust mass is surviving in Cas~A, with part of the destruction occurring in the shocked and unshocked ISM. We note that \cite{Bocchio2016} did not consider clumps in the ejecta which will (at least partly) reduce the ejection of dust grains into the shocked ISM. 

\cite{Slavin2020} conducted 2D hydrodynamical simulations of an expanding SNR within the first 8000$\,$yr and extrapolated the conditions in a 1D approximation up to $0.1\,$Myr. They calculated the sputtering of dust for the clumpy ejecta of Cas~A and found a silicate survival fraction between $0.3\,$per~cent ($a_\text{peak}=40\,$nm) and $22\,$per~cent  ($625\,$nm). Similar to the study of \cite{Bocchio2016} the
dust material is subject to a significant destruction after the grains have left the shocked ejecta and passed the forward shock. Due to the long evolution time of 8000$\,$yr and an accordingly huge ejecta (${\sim}20\,$pc) and domain size, the ejecta clumps were only marginally resolved and the modelling of clump disruption is thus only partially realized. They omitted grain-grain collisions and magnetic field effects on both the gas and the dust which would cause even lower dust survival rates.

\cite{Fry2020} used 1D MHD simulations to study the evolution of dust grains that are formed in clumps in an expanding SNR and which have to pass the reverse shock. Though  sputtering is considered in their simulations, they do not report dust survival rates. Interestingly, they find  dust grains which can be trapped at the interface between the SN ejecta and shocked ISM or reflected back into the ejecta and forth multiple times, preventing the dust grains from overtaking the forward shock and escaping into the ISM. However, it is unclear whether this pinball effect is a result of the intrinsic magnetic field structure of the chosen model.

To summarize the comparison with previous studies: Our present work is the only one using MHD simulations of an evolving SNR, modelling clumps and the clump disruption at high resolution, while grain-grain collisions and the influence of magnetic fields on the dust dynamics are taken into account. However, our study shows deficits in the long-term processing of the dust grains. We are going to discuss these points in the following section.

\subsection{Model deficits and potential for improvements}

In our simulations, the gradual disruption of a clump is followed from the first contact with the reverse shock up to a period of three cloud-crushing times. During this time interval, the properties of the clump and the surrounding gas are not affected by the expansion of the ejecta due to the SNR evolution. This static approximation is reasonable because the cloud-crushing simulations last less than 15$\,$per~cent of the remnant's age in any case. If the expansion of the SNR ejecta and the clump disruption were considered simultaneously, 
slightly less dust destruction than predicted in our multi-step model is to be expected due to the gradual decrease of the gas density (Fig.~\ref{AB_initial}). 

On the other hand, the dust destruction in our simulations can only take place during the three cloud-crushing timescales, but not beyond. At first glance, this is a limitation of our model. However, following \cite{Silvia2010}, dust destruction in a cooled post-shock gas 
is most effective in the first three cloud-crushing times. Due to the gradual decrease of the gas density with increasing remnant age, the effect of continued dust destruction beyond three cloud-crushing times becomes less and less important. Considering the dust destruction over longer periods of time would increase the dust destruction in young remnants, for which the dust destruction is already near total, and to a lesser extent in old remnants. Overall, a somewhat increased level of cumulative dust destruction would be expected. Thus, with the static approximation during the cloud-crushing simulations and the dust destruction during the three cloud-crushing times only, there are two opposing effects. We expect some degree of cancellation between these two effects and that the combination of the effects will not have a significant impact on our results.

Other aspects to be discussed are the evolution of the clumps over time and the clump morphology. We assumed a fixed clump mass and a fixed pressure equilibrium between clump and interclump medium, and as a result the clump expands with the expansion of the SNR (Section~\ref{Berlin}). This is a plausible assumption, but it is accompanied by other uncertainties. First, it is unclear when and how exactly clumps form in SN ejecta and SNRs. This depends, e.g., on the formation history of the clumps, since they can be built up from inhomogeneities in the SN explosion or by instabilities occurring during the ejecta or remnant expansion. Furthermore, the clump frequency does not have to be proportional to the ejected gas density, but can follow a more complex relation, which has an influence on when and how many clumps are hit by the reverse shock, significantly determining the dust survival fraction. It is also conceivable that the clump mass increases during the initial phase, since the clump has to form first. More precise modelling is not reasonable unless future observations can further constrain initial phase of the clump formation process. We have made as few assumptions as possible about the evolution of clumps in our model to simulate a realistic scenario.

Regarding the 3D morphology of the clumps, it is obvious that a perfectly spherical clump shape (or circular disc) is merely an assumption to keep the model approach simple. Observations by \textsl{Hubble} and recently by JWST reveal a complex structure of the Cas~A ejecta (e.g. \citealt{Ennis2006, Smith2009, Milisavljevic2013, Milisavljevic2024}), with highly-elongated  filaments, fractal density enhancements, and asymmetric structures. Although we showed in \cite{Kirchschlager2023} that doubling the radius of a spherical clump has only a small effect on the dust survival fraction, the expected outcome for much more complex structures is not predictable. In particular, the shock velocity will not be uniform at a certain time and could also point in different directions, significantly affecting the amount of shocked gas mass and the destroyed dust fractions. We note that in our study the SN expansion was modelled in a smooth and clump-free remnant. For a remnant in which a large fraction of the total mass is contained in clumps, the evolution could be significantly different.

Uncertainties also include the gas-to-dust mass ratio, the total ejecta mass and the dust mass in Cas~A. Given a gas-to-dust mass ratio of $10$ in the clumps (\citealt{Priestley2019a}) and an ejecta mass of $2.2\,\text{M}_\odot$ (\citealt{Willingale2002}), the unprocessed dust mass is less than ${\sim}0.2\,\text{M}_\odot$ which is in contradiction to the dust masses derived from observations ($0.5-1.1\,\text{M}_\odot$; e.g. \citealt{Bevan2017,DeLooze2017, NiculescuDuvaz2021, Priestley2022b}). Recent studies pointed to higher ejecta masses ($3-4\,\text{M}_\odot$; \citealt{Hwang2012, Lee2014, Orlando2016, Arias2018}) and lower gas-to-dust mass ratios (in the order of ${\sim}1$; \citealt{Priestley2022b}) which would affect the dust survival fractions, however, better constraints from observations are needed. If the reverse shock forms later on, dust destruction will be delayed and thus alter the observable dust mass at the age of ${\sim}350\,$yr (today). Due to cooling and density dilution as consequence of the SNR expansion, an increased dust survival rate can be expected over the remnant evolution. Moreover, the SN explosion energy and the density of the surrounding CSM are not well constrained; in combination with the ejecta mass they define the index $n$ (Section~\ref{Prag}) which significantly affects the remnant evolution and which is adjusted to well reproduce the observables of Cas~A like the remnant extension and the reverse shock position. An overall model reconciling all the various properties does not yet exist and is beyond the scope of this study, however, we have been able to reproduce significant ejecta properties which are either inferred from observations or which are consistent with other studies for comparability. 

The goal of our study is to determine the evolution of dust in a SNR in which a reverse shock occurs. The reverse shock characteristics depend crucially on the SN explosion energy (of the order of $10 ^{51}\,$ergs), the ejecta mass (0 -- a few M$_\odot$) and the (mean) gas density of the surrounding CSM (broad range between ${\lesssim}1$ and ${\gtrsim}10^3\,$cm$^{-3}$), which are particular for each SNR. The example of the Crab Nebula shows that it can even happen that no reverse shock occurs 1000 years after the SN explosion and therefore no dust processing has taken place so far. In the present study, we focused on Cas~A, which has been extensively studied and for which many ejecta properties are relatively well known. However, we would like to emphasize that the different conditions in and around other SNRs will lead to a different course for the reverse shock and therefore to a different dust destruction fraction. Besides the SN explosion energy, ejecta mass and CSM density, the time and position of the formation of the reverse shock and its velocity (and thus the dust processing) will depend on further effects and conditions like the cooling efficiency or the temperature of the CSM. A detailed examination of the evolution of the dust survival in other SNRs has to be carried out separately for each system which is far beyond the scope of the present study. Because of the complexity of the formation process of the reverse shock, we refrain from attempting a rough extrapolation from the results of the modelling of Cas~A to other SNRs in order to prevent a misjudgement.

%

The last point we want to bring up for discussion is the potential evolution of dust in the SNR. In our model we assume the dust already exists when the reverse shock hits the first clumps. This is in agreement with dust formation theories, which predict that the dust will be fully formed within a few hundred or thousand days after the SN explosion (e.g. \citealt{Nozawa2003}). However, observations of the infrared excess and the red-blue asymmetry of emission lines indicate that the dust mass gradually increases over decades (\citealt{Gall2014, Wesson2021, NiculescuDuvaz2022}). In this case, the reverse shock in very young SNRs will destroy less dust than our model predicts, since, trivially, the dust has not yet formed. Similar to the clump occurrence rate in the ejecta, which could follow a more complex relation, it is also conceivable that not every clump in the remnant has the same dust properties, such as the same dust grain sizes and dust materials. The latter is also plausible since the elemental abundances like in Cas~A are not uniformly distributed (e.g. \citealt{Arendt2014}). As shown above but also in previous studies, the dust grain sizes are an important factor for the resulting dust survival rates. Except for accretion and ion trapping of dusty gas on dust grain seeds, we also did not account for dust re-formation in the post-shock gas, e.g. in dense shells between forward and reverse shock or in a cool dense shell formed by an ejecta/CSM collision (\citealt{Kotak2009,Mauerhan2017, Sarangi2022}), which would lead to higher surviving dust mass fractions, implying that SNe could contribute more material to the galactic dust budgets.

In summary, the consideration of more complex distributions of grain sizes, dust materials, clump structures and clump occurrence rates in the SNR is beyond the scope of this study.
 
\section{Conclusions}

This is to our knowledge the first study that investigates the interaction of ejecta clumps with the reverse shock in an evolving SNR in which dust grains can be destroyed by the combined effects of sputtering and grain-grain collisions. The dust survival fractions were determined in three consecutive simulation steps for the SNR Cas~A: In the first step the expansion of a smooth and clump-free SNR was modelled for $1000\,$yr using a 1D MHD simulation. The resulting shock and ejecta properties were then used to conduct 2D MHD simulations to model the impact of the reverse shock on a single ejecta clump with time-varying initial conditions. In the last step the dust dynamics and dust destruction fractions in the shocked clump are determined by running post-processing simulations of the cloud-crushing results. Knowing the shocked ejecta gas masses and the dust destruction fractions at each age of the evolving remnant finally allowed us to determine the overall dust survival rate during the remnant evolution. 

We found total destruction of the dust in clumps that are shocked within the first $200\,$yr. Between 200 and $1000\,$yr the survival fractions per clump increase continuously, specific results depend on the initial grain sizes. At remnant ages of ${\sim}1000\,$yr the survival fractions per clump encountering the reverse shock for the first time converge to ${\sim}100\,$per cent for all grain sizes. Sputtering is the dominant factor for most grain sizes and remnant ages, however, grain-grain collisions additionally reduce the survival fraction and become most effective for $100-1000\,$nm grains.

Accordingly, the dust mass integrated along the entire SNR shows a steep decline in the first $200\,$yr before the dust destruction weakens and the mass converges to its final value. The dust mass of the smallest grains ($1\,$nm) shows the lowest (17 per~cent) and the largest grains the highest survival fraction (28 per~cent). We find dust destruction after $1000\,$yr to be negligible in Cas~A. 

Finally, we would like to emphasize that, for the question of dust survival in an SNR, the temporal evolution of the ejecta must also be taken into account in addition to the previously known factors of particle size, clump density and magnetic field strength.
\label{Dublin}

 
 \section*{Acknowledgements}
FK, NSS and IDL have received funding from the European Research Council (ERC) under the European Union’s Horizon 2020 research and innovation programme DustOrigin (ERC-2019-StG-851622). NSS acknowledges the support from the Flemish Fund for Scientific Research (FWO-Vlaanderen) in the form of a postdoctoral fellowship (1290123N). FDP acknowledges the support of a consolidated grant (ST/K00926/1) from the UK Science and Technology Facilities Council (STFC). Simulations were performed using the data intensive {\textsc{Peta4-Skylake}} and {\textsc{Peta4-Icelake}} service at Cambridge, supported through DiRAC project ACSP190 (SNDUST) using the Cambridge Service for Data Driven Discovery (CSD3), part of which is operated by the University of Cambridge Research Computing on behalf of the STFC DiRAC HPC Facility (\href{www.dirac.ac.uk}{www.dirac.ac.uk}). The DiRAC component of CSD3 was funded by BEIS capital funding via STFC capital grants ST/P002307/1 and ST/R002452/1 and STFC operations grant ST/R00689X/1. DiRAC is part of the U.K. National \mbox{e-Infrastructure.}\vspace*{-0.3cm}

 \section*{Data availability}
 The data underlying this article will be shared upon reasonable request to the corresponding author.



{\footnotesize
  \bibliography{Literature}
}




   \appendix

    \section{Reverse shock velocity}
   \label{app4}   
       \begin{figure}
    \resizebox{\hsize}{!}{
 \includegraphics[trim=2.35cm 15.05cm 2.3cm 1.8cm, clip=true,  page=1]{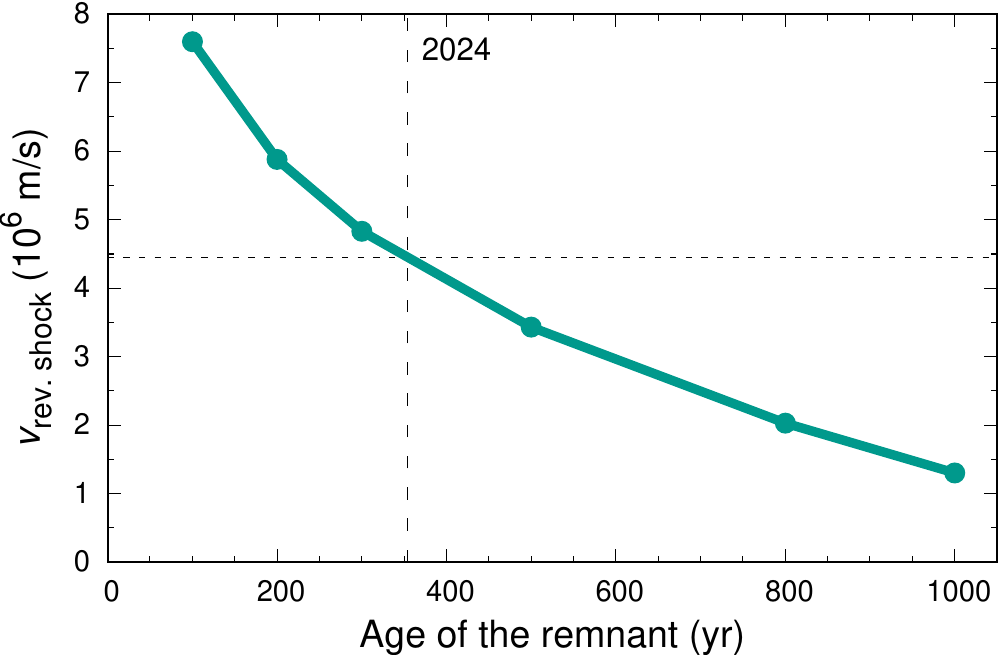}}\vspace*{-0.15cm} 
   \caption{\label{Revshockvelocity}Reverse shock velocity in the rest frame of the explosion centre in outbound direction. For comparison, Fig.~\ref{Arepo_result} (second row) shows the reverse shock velocity in the frame of the moving ejecta. The vertical line indicates the year 2024.}
\end{figure}
In Fig.~\ref{Revshockvelocity} the reverse shock velocity is shown within the first $1000\,$yr after the SN explosion in the rest frame of the explosion centre.


\label{lastpage}
\end{document}